\documentclass[preprint,aps]{revtex4}

\usepackage{graphicx}
\usepackage{dcolumn}

\begin{document}

\preprint{Lebed-PRL}

\title{Cooper Pairs with Broken Parity and Time-Reversal
Symmetries in D-wave Superconductors}

\author{A.G. Lebed$^*$}

 \affiliation{Department of Physics, 
University of Arizona,
1118 E. 4-th Street, Tucson, AZ 
85721, USA}

\begin{abstract}
Paramagnetic effects are shown to result in the appearance 
of a triplet component of order parameter in a
vortex phase of a d-wave superconductor in the
absence of impurities.
This component, which breaks  both parity and time-reversal 
symmetries of Cooper pairs, is expected to be of the order of unity 
in a number of modern superconductors such as organic, 
high-T$_c$, 
and some others.
A generic phase diagram of such type-IV superconductors, 
which are singlet ones at $H=0$ and characterized by 
singlet-triplet mixed Copper pairs with broken time-reversal 
symmetry in a vortex phase,
is discussed.
\end{abstract}

\maketitle
  
\pagebreak
 
It is well known [1,2] that type-II superconductors, where
superconductivity survives at high magnetic fields, 
$H_{c1} (T) < H < H_{c2} (T)$, as Abrikosov vortex phase [3,2,1], 
are subdivided into two main
classes.
They are superconducting alloys (or dirty superconductors) [1,2]
and relatively clean materials, where type-II superconductivity is
due to anisotropy of their quasi-particles spectra and relatively
heavy masses of quasi-particles [4].
The latter compounds are currently the most interesting and 
perspective superconducting materials, including organic [5],
heavy fermions [6], high-T$_c$ [7], MgB$_2$ [8], and some other 
superconductors.  

Superconducting orbital order parameter, $\Delta({\bf r_1},{\bf r_2})$, 
corresponding to pairing of two electrons in Cooper pair, can 
usually be expressed as 
$\Delta({\bf r}_1, {\bf r}_2) =  \Delta({ \bf R})  \Delta ({\bf r})$ 
[9,10], where external order parameter, $\Delta({\bf R})$, is related 
to  motion of a center of mass of Cooper pair, ${\bf R}=({\bf r_1}+{\bf r_2})/2$, 
whereas internal order parameter, $\Delta ({\bf r})$, describes 
relative motion of electrons in 
Cooper pair,
${\bf r}={\bf r_1}-{\bf r_2}$.
In this context, type-II superconductors in their vortex phases
are characterized by broken symmetries of $\Delta({ \bf R} )$, 
corresponding to vortices and Meissner 
currents.
Other important issues are symmetries of internal orbital
order parameter, $\Delta ({\bf r})$, and related spin part 
of order parameter, 
$\Delta (\sigma_1, \sigma_2)$.
To satisfy Fermi statistics, in case of singlet superconductivity
(where the total spin of Cooper pair $|{\bf S}|= 0$), internal order
parameter, $\Delta ({\bf r})$, has to be an even function of 
coordinate ${\bf r}$, whereas, in  case of triplet superconductivity 
(where $|{\bf S}| = 1$), $\Delta ({\bf r})$ has to be an odd function 
of ${\bf r}$.
In accordance with symmetry properties of $ \Delta ({\bf r})$ 
(or its Fourier component, $\hat{ \Delta} ({\bf k})$),
superconductors are subdivided into conventional ones [1,2]
(where superconductivity is described by BCS s-wave singlet 
pairing) and unconventional ones [9,10]  (where symmetry of 
$\hat{ \Delta} ({\bf k})$ is lower than 
underlying symmetry of 
crystalline lattice).
At present time, unconventional d-wave singlet superconductivity 
is firmly established in high-T$_c$ [11] and some organic 
materials.
There exist also several strong candidates for unconventional 
triplet superconducting pairing such as Sr$_2$RuO$_4$ [12], 
(TMTSF)$_2$X [13], ferromagnetic [14], and heavy fermion [9,10,15]
superconductors.
It is a common belief that magnetic field does not
change internal order parameters, $\hat{ \Delta} ({\bf k})$ and 
$\Delta (\sigma_1, \sigma_2)$, in conventional [1,2] and 
unconventional [9,10] 
type-II superconductors. 
Moreover, although Meissner currents break time-reversal 
symmetry of $\Delta(R)$, internal  order parameters, 
$\hat{ \Delta} ({\bf k})$ and $\Delta (\sigma_1, \sigma_2)$, 
are believed to 
preserve $t \rightarrow -t$
symmetry. 

The goal of our Letter is to demonstrate that there have to exist
type-IV superconductors [16], which exhibit qualitatively different
magnetic properties. 
More precisely, we suggest and prove the following theorem: 
each singlet type-II superconductor in the absence of impurities is 
actually type-IV superconductor with broken  ${\bf k} \rightarrow {\bf -k}$ 
and $t \rightarrow -t$ symmetries of internal Cooper pairs wave 
functions in vortex phase, provided that effective constant of triplet 
pairing
is not exactly zero, 
$g_t \neq 0$.
We show that the above mentioned theorem is an inherent property 
of singlet superconductivity and is due to careful account for 
paramagnetic spin-splitting effects in vortex phase, which 
have been treated so far only for $g_t = 0$ [17,1].

We define type-IV superconductivity as singlet superconductivity 
at $H=0$ which exhibits  broken symmetries of internal Cooper 
pairs wave functions, $\hat{ \Delta} ({\bf k})$ and $\Delta
(\sigma_1, \sigma_2)$, 
in vortex phase.
In our particular case, internal order parameter 
in vortex phase is shown to be a mixture of a singlet d-wave component, 
$ \hat{ \Delta_s} ({\bf k})$, with a triplet component, 
$i \  \hat{ \Delta_t} ({\bf k})$.
Note that this order parameter breaks not only parity symmetry, 
${\bf k} \rightarrow -{\bf k}$, but also a time-reversal symmetry,
$t \rightarrow -t$, due to an imaginary 
coefficient $i$.
Below, we demonstare that the effects of singlet-triplet
coexistence  are of the order of unity in a number of  modern clean 
type-II superconductors, where the orbital upper critical fields 
are of the order of paramagnetic 
limiting fields, $\mu_B H_{c2} (0)
\sim T_c$ [18] (see Table 1).
It is important that the suggested theorem  is very general: 
it is valid even for simplest spin independent electron-electron 
interactions for both attractive and repulsive 
interactions in triplet 
channel. 
As discussed below, this theorem is based on symmetry arguments
and is a consequence of broken spin symmetry (due to paramagnetic 
effects) and broken translational invariancy of $\Delta(R)$ 
(due to the existence of 
vortices).

Here, we discuss how paramagnetic effects result 
in the appearance of a triplet component in 
vortex phase using qualitative arguments.
We recall that, in vortex phase, external order parameter,
$\Delta(R)$, is a function of ${\bf R}$ on scale of $\xi$,
where $\xi$ is a coherence length [1-4].
Therefore,  $\Delta(R)$ corresponds to superconducting pairing 
of electrons with total non-zero momenta of Cooper pairs of 
the order of
$|{\bf q}| \sim \hbar / \xi$.
As seen from Fig.1, a probability of pairing for 
electrons with spin up and spin down, $|\Delta(+, -)|^2$, 
is different from that for electrons with spin down and spin up, 
$|\Delta(-, +)|^2$, if 
${\bf q} \neq 0$.
Therefore, singlet superconductivity, which is characterized
by spin order parameter $\Delta (+, -) = - 
\Delta (-, +)$ [9,10], has to be mixed with a triplet component,
characterized by spin order parameter $\Delta (+, -) =  \Delta (-, +)$ 
(i.e., $|{\bf S}| =1$ and
$S_y =0$) (see Fig.1).
   
Below, we quantitatively  describe superconducting pairing 
with internal order parameter, exhibiting broken inversion and
 time-reversal symmetries, in d-wave singlet superconductor
 with layered electron spectrum,  
\begin{equation}
\epsilon_0({\bf k})= (k^2_x + k^2_y) / 2m + 2 t_{\perp} \cos(k_z d) \ , \ \ \ 
\epsilon_F = m v^2_F / 2 \ ,
\end{equation}
in a magnetic field:
\begin{equation}
{\bf H} = (0,H,0) \  ,  \ \ \ \ {\bf A} = (0,0,-Hx) \ .
\end{equation}
In case, where electron-electron interactions do not 
depend on electron spins, the total Hamiltonian of electron 
system can be written in the form:
\begin{eqnarray}
 &&\hat{H} = \hat{H}_0 + \hat{H}_{int} \ , \ \ \ \ \ 
\hat{H}_0 = \sum_{{\bf k} ,  \sigma}  
\epsilon_{\sigma} ({\bf k})  \ a^{+}_{\sigma} ({\bf k}) \ a_{\sigma} ({\bf k}) \ ,
\nonumber\\ 
&&\hat{H}_{int} = \frac{1}{2} \sum_{{\bf q} ,  \sigma}  \sum_{{\bf k} , {\bf k_1}}  \ 
V({\bf k} , {\bf k_1}) \ 
a^{+}_{\sigma} ({\bf k} + \frac{{\bf q}}{2})  \ a^{+}_{-\sigma} ({-\bf k} + \frac{{\bf q}} {2})
\ a_{-\sigma} ({-\bf k_1} + \frac{{\bf q}}{2}) \  a_{\sigma} ({\bf k_1} + \frac{{\bf q}}{2}) \ ,
\end{eqnarray} 
where  $\epsilon_{\sigma} ({\bf k}) = \epsilon_0({\bf k}) - 
\sigma \mu_B H$ ($\sigma = \pm 1$), $a^{+}_{\sigma} ({\bf k})$ and 
$a_{\sigma} ({\bf k})$
are electron creation and annihilation operators. 
As usually [9,10], electron-electron interactions are subdivided 
into singlet and triplet ones:
\begin{eqnarray}
&&V({\bf k}, {\bf k_1}) = V_s({\bf k}, {\bf k_1}) + V_t({\bf k}, {\bf k_1}) \ , \ \ 
V_s({\bf k}, {\bf k_1}) = V_s(-{\bf k}, {\bf k_1}) = V_s({\bf k}, - {\bf k_1}) \ ,
\nonumber\\ 
&&V_t({\bf k}, {\bf k_1}) = - V_t(-{\bf k}, {\bf k_1}) = - V_t({\bf k}, - {\bf k_1})    \ .
\end{eqnarray} 
We define normal and Gorkov (anomalous) finite temperature Green 
functions, 
\begin{eqnarray}
&&G_{\sigma, \sigma}({\bf k}, {\bf k_1}; \tau)  = 
- < T_{\tau} a_{\sigma}({\bf k}, \tau) a^+_{\sigma}({\bf k_1}, 0 > , \ \  
F_{\sigma, -\sigma}({\bf k}, {\bf k_1}; \tau)  = 
< T_{\tau} a_{\sigma}({\bf k}, \tau) a_{-\sigma}({-\bf k_1}, 0) > \ ,
\nonumber\\ 
&&F^+_{\sigma, -\sigma}({\bf k}, {\bf k_1}; \tau)  = 
< T_{\tau} a^+_{\sigma}({-\bf k}, \tau) a^+_{-\sigma}({\bf k_1}, 0)  >   \ ,
\end{eqnarray} 
as well as singlet and triplet superconducting order parameters,
\begin{eqnarray}
&&\Delta_s ({\bf k}, {\bf q}) = - \frac{1}{2} \sum_{ {\bf k}_1}  V_s({\bf k},{\bf k_1}) 
T \sum_{\omega_n} 
[F_{+, -}(i \omega_n ; {\bf k}_1+ \frac{{\bf q}}{2}, {\bf k_1}- \frac{{\bf q}}{2}) 
- F_{-, +}(i \omega_n ; {\bf k}_1+\frac{{\bf q}}{2}, {\bf k_1}-\frac{{\bf q}}{2})]  \ , 
\nonumber\\
&&\Delta_t ({\bf k}, {\bf q}) = - \frac{1}{2} \sum_{ {\bf k}_1}  V_t({\bf k},{\bf k_1}) 
T \sum_{\omega_n} 
[F_{+, -}(i \omega_n ; {\bf k}_1+\frac{{\bf q}}{2}, {\bf k_1}-\frac{{\bf q}}{2})
+ F_{-, +}(i \omega_n ; {\bf k}_1+\frac{{\bf q}}{2}, {\bf k_1}-\frac{{\bf q}}{2})]  \ , 
\end{eqnarray} 
by standard ways [19,20,9,10].

The goal of our Letter is to consider a phase transition line 
between metallic and singlet-triplet mixed superconducting 
phases in Ginzburg-Landau (GL) region, $(T_c-T)/ T_c \ll 1$ [1-4], 
where $T_c$ is a transition temperature between metallic  
state and d-wave singlet  
phase at $H=0$.
For this purpose, we linearize Gorkov equations [9,10,19] with 
respect to order parameters (6) and obtain the following system 
of linear equations [21]:
\begin{eqnarray}
&&\Delta_s ({\bf k}, {\bf q}) = - \frac{1}{2} \sum_{ {\bf k}_1}  V_s ({\bf k},{\bf k_1}) 
T \sum_{\omega_n} [ \Delta_s ({\bf k_1}, {\bf q}) \ S 
+ \Delta_t ({\bf k_1}, {\bf q}) \ D]  \ , 
\nonumber\\
&&\Delta_t ({\bf k}, {\bf q}) = - \frac{1}{2} \sum_{ {\bf k}_1}  V_t ({\bf k},{\bf k_1}) 
T \sum_{\omega_n} [ \Delta_t ({\bf k_1}, {\bf q}) \ S 
+ \Delta_s ({\bf k_1}, {\bf q}) \ D]   \ , 
\nonumber\\
&&S   =
G^0_{+} (i \omega_n, {\bf k}_1+ \frac{{\bf q}}{2})
G^0_{-} (-i \omega_n, -{\bf k}_1+ \frac{{\bf q}}{2}) +
G^0_{-} (i \omega_n, {\bf k}_1+ \frac{{\bf q}}{2})
G^0_{+} (-i \omega_n, -{\bf k}_1+ \frac{{\bf q}}{2})   \ , 
\nonumber\\
&&D  =   
G^0_{+} (i \omega_n, {\bf k}_1+ \frac{{\bf q}}{2})
G^0_{-} (-i \omega_n, -{\bf k}_1+ \frac{{\bf q}}{2}) -
G^0_{-} (i \omega_n, {\bf k}_1+ \frac{{\bf q}}{2})
G^0_{+} (-i \omega_n, -{\bf k}_1+ \frac{{\bf q}}{2}) \ ,
\end{eqnarray} 
where $G^0_{\sigma} (i \omega_n, {\bf k}) = 
1 / (i \omega_n - \epsilon_{\sigma}({\bf k}))$ 
is Green function of a free electron in the presence
of paramagnetic effects and $\omega_n$ is Matsubara
frequency [20]. [Note that common Eqs.(7)
directly demonstrate singlet-triplet coexistence
effects in vortex phase, where ${\bf q} \neq 0$ 
(see Fig.1)].

Below, we consider in detail an important example, 
coexistence of singlet $d_{x^2-y^2}$-wave [21] and
triplet $p_{x}$-wave order parameters, which
corresponds to the following matrix elements of
electron-electron interactions:
\begin{equation}
\label{evp}
\left(
    \begin{array}{c}
        V_s({\bf k},{\bf k}_1) \\
        V_t({\bf k},{\bf k}_1) 
    \end{array}
\right)
= - \frac{4 \pi }{v_F}
\left(
    \begin{array}{c}
       g_s \   \cos 2\phi \  \cos 2\phi_1 \\
       g_t \ \cos (\phi - \phi_1)
    \end{array}
\right)
, \ \ g_s > 0, 
\ g_s > g_t \ ,
\end{equation} 
where  $\phi$ and $\phi_1$ are polar angles corresponding to 
momenta ${\bf k}$ and ${\bf k_1}$, respectively.
[Note that inequalities $g_s > 0$ and $g_s > g_t$ guarantee 
that singlet $d_{x^2-y^2}$-wave phase is a 
ground state at $H=0$ 
and $T<T_c$].
After substitution of Eqs.(8) in Eqs.(7), we represent 
order parameters as follows,
$\Delta_s({\bf k}, {\bf q}) = \sqrt{2} \cos 2\phi \ \Delta_s({\bf q})$
and  $\Delta_t({\bf k}, {\bf q}) = \sqrt{2} \cos \phi \ \Delta_t({\bf q})$,
and rewrite Eqs.(7) in a matrix form:
\begin{equation}
\label{evp}
\left(
    \begin{array}{cc}
       A_{ss} ({\bf q}) &   A_{st} ({\bf q}) \\
       A_{ts} ({\bf q}) &  A_{tt} ({\bf q})

    \end{array}
\right)
\left(
    \begin{array}{c}
        \Delta_s({\bf q}) \\
        \Delta_t( {\bf q})
    \end{array}
\right)
=
\left(
    \begin{array}{c}
       \Delta_s({\bf q}) / g_s\\
       \Delta_t( {\bf q}) / g_t
    \end{array}
\right)
\ .
\end{equation}

We calculate matrix $\hat{A}({\bf q})$ at $q_y =0$ 
in GL region [3,4,22,9] which corresponds to its expansion 
as power series in small parameters,
$v_F q_x /T_c \ll 1$ and  $t_{\perp} d q_z /T_c \ll 1$.
As a result, we obtain:
\begin{equation}
\hat{A}
=
\label{evp}
\left(
    \begin{array}{cc}
      (2 \pi T) \sum^{\Omega}_{\omega_n >0} 
\biggl[ \frac{1}{\omega_n} - \frac{1}{8 \omega^3_n} (v^2_Fq^2_x + 
4 t^2_{\perp} q^2_z d^2) \biggl]  \  \ , 
       & \  -  \mu_B H v_F q_x (\pi T_c) \sum^{\infty}_{\omega_n >0} \frac{1}{\omega^3_n} \\
        -  \mu_B H v_F q_x (\pi T_c) \sum^{\infty}_{\omega_n >0} \frac{1}{\omega^3_n} \ \ ,
& (2 \pi T) \sum^{\Omega}_{\omega_n >0} 
\biggl[ \frac{1}{\omega_n} - \frac{1}{8 \omega^3_n} (3 v^2_Fq^2_x /2 + 
4 t^2_{\perp} q^2_z d^2) \biggl] 
    \end{array}
\right) \ ,
\end{equation}
with $\Omega$ being a cut-off energy.
Magnetic field (2) is introduced in Eqs.(9),(10) by means of 
a standard quasi-classical approximation [23,22,4,3],
$ q_x \rightarrow - i ( d / dx) , \ q_z / 2 \rightarrow eA_z / c = e H x / c$,
which leads to the following matrix GL  equations extended to the case of 
triplet-singlet coexistence:
\begin{equation}
\label{evp}
\left(
    \begin{array}{cc}
       \tau + \xi^2_{\parallel} \biggl( \frac{d^2}{dx^2}  \biggl)
       - \frac{(2\pi \xi_{\perp})^2}{\phi^2_0} H^2 
       x^2  \ , 
       & \ i \frac{ \sqrt{7 \zeta (3)} }{\sqrt{2} \pi} 
\biggl( \frac{\mu_B H}{T_c} \biggl)  \xi_{\parallel}  \biggl(  \frac{d}{dx} \biggl)  \\
 \ i \frac{ \sqrt{7 \zeta (3)} }{\sqrt{2} \pi} 
\biggl( \frac{\mu_B H}{T_c} \biggl)  \xi_{\parallel}  \biggl(  \frac{d}{dx} \biggl) \ ,            
       &  \  \tau+ \frac{g_t - g_s}{g_t g_s} + \frac{3}{2} \xi^2_{\parallel} 
       \biggl( \frac{d^2}{dx^2} \biggl) - 
\frac{(2\pi \xi_{\perp})^2}{\phi^2_0} H^2 x^2
    \end{array}
\right)
\left(
    \begin{array}{c}
        \Delta_s(x) \\
        \Delta_t(x)
    \end{array}
\right)
= 0 \ ,
\end{equation}
where $\tau = (T_c -T)/T_c \ll 1$, $\xi_{\parallel} = \sqrt{7 \zeta (3)} v_F / 
4 \sqrt{2} \pi T_c$ and $\xi_{\perp} = \sqrt{7 \zeta (3)} t_{\perp} d / 
2 \sqrt{2} \pi T_c$ are GL coherence lengths [4,9], $\zeta(3) \simeq 1.2$ 
is zeta Riemann function, $\phi_0$ is a flux quantum, $x$ is coordinate
of a center of mass of Cooper pair.
In typical case, where  $g_s - g_t \sim g_s$, Eqs. (11) have the following
solutions:
\begin{equation}
\label{evp}
\left(
    \begin{array}{c}
        \Delta_s(x)  \\
        \Delta_t(x) 
    \end{array}
\right)
= 
\left(
    \begin{array}{c}
      \exp \biggl(- \frac{\tau x^2}{ 2 \xi^2_{\parallel}} \biggl) \\
       i \  \sqrt{\tau}  \frac{ \sqrt{7 \zeta (3)} }{ \pi} 
\biggl(  \frac{ g_t g_s }{ g_t - g_s }  \biggl)
\biggl( \frac{\mu_B H}{T_c} \biggl) 
\biggl( \frac{ \sqrt{\tau} x }{\sqrt{2} \xi_{\parallel}} \biggl)
 \exp \biggl(- \frac{\tau x^2}{ 2 \xi^2_{\parallel}} \biggl)
    \end{array}
\right) \ .
\end{equation} 

Eqs.(11),(12) are the main results of our Letter. 
They extend GL differential equation [1-4,22,9,10] and 
its Abrikosov  solution for superconducting nucleus,
$\exp (- \tau x^2/2 \xi^2_{\parallel})$ 
[3,1,2], to the case
$g_t \neq 0$. 
Eqs.(11),(12) directly demonstrate that, in vortex phase,
 singlet order parameter always coexists with  triplet one, 
 characterized by $|{\bf S}|=1$  and $S_y=0$ 
 (${\bf H} \parallel {\bf y}$), for arbitrary sign of effective 
 triplet coupling constant $g_t$.
It is important that triplet  component (12) breaks not
only parity  but also time-reversal symmetry due to
the existence of non-diagonal matrix elements, proportional
to $i H$, in Eqs.(11).
 Indeed, $\Delta^*_t(x)  \neq \Delta_t(x)$ in Eq.(12),  which 
 indicates that $t \rightarrow -t$ symmetry is broken and,
 thus, Cooper pairs possess some 
 internal magnetic 
moments [9,10].

To summarize, the main message of the Letter is that Cooper pairs 
cannot be considered as unchanged elementary particles 
in vortex phases of modern strongly correlated type-II superconductors, 
where $\mu_B H_{c2}(0) \sim T_c$
and $|g_s| \sim |g_t|$.
Indeed triplet-singlet components ratio in Eqs. (12) at 
$x=\sqrt{2} \xi_{\parallel} / \sqrt{\tau}$ and low
temperatures, $\tau \sim 1$, can be estimated as 
$R = \Delta_t / \Delta_s \sim  i \ 
[\mu_B H_{c2}(0) / T_c]$ (see Table 1).
The appearance of a triplet component, breaking time-reversal
symmetry, has to change all qualitative features of vortex
phases in d-wave superconductors.
These include the appearance of non-zero internal magnetic
moments of Cooper pairs, possible unusual topology of superconducting 
vortices, the appearance of spin-wave excitations, the disappearance 
of quasi-particles near zeros of $d_{x^2-y^2}$-wave gap, $\cos 2 \phi
=0$, unusual spin susceptibility, and other non-trivial phenomena to be 
studied in
the future.
We suggest that, in clean type-II superconductors, there  
exist the forth critical fields, $H_{c4}(T)$, corresponding to phase
transitions (or crossovers) between Abrikosov vortex phases and some 
exotic vortex phases and call such materials type-IV 
superconductors. 
 In conclusion, we point out that singlet-triplet mixing effects were earlier 
 studied in He$^3$ [24], Larkin-Ovchinnikov-Fulde-Ferrell phase [25,26], 
 for surface superconductivity [27], and in superconductors without
 inversion symmetry [9,28,29]. 

The author is thankful to N.N. Lebed (Bagmet), P.M. Chaikin, and 
V.M. Pudalov for numerous and useful discussions.

$^*$Also Landau Institute for Theoretical Physics, 
2 Kosygina Street, Moscow, Russia.

\pagebreak

\begin{figure}[h]
\includegraphics[width=7.6in,clip]{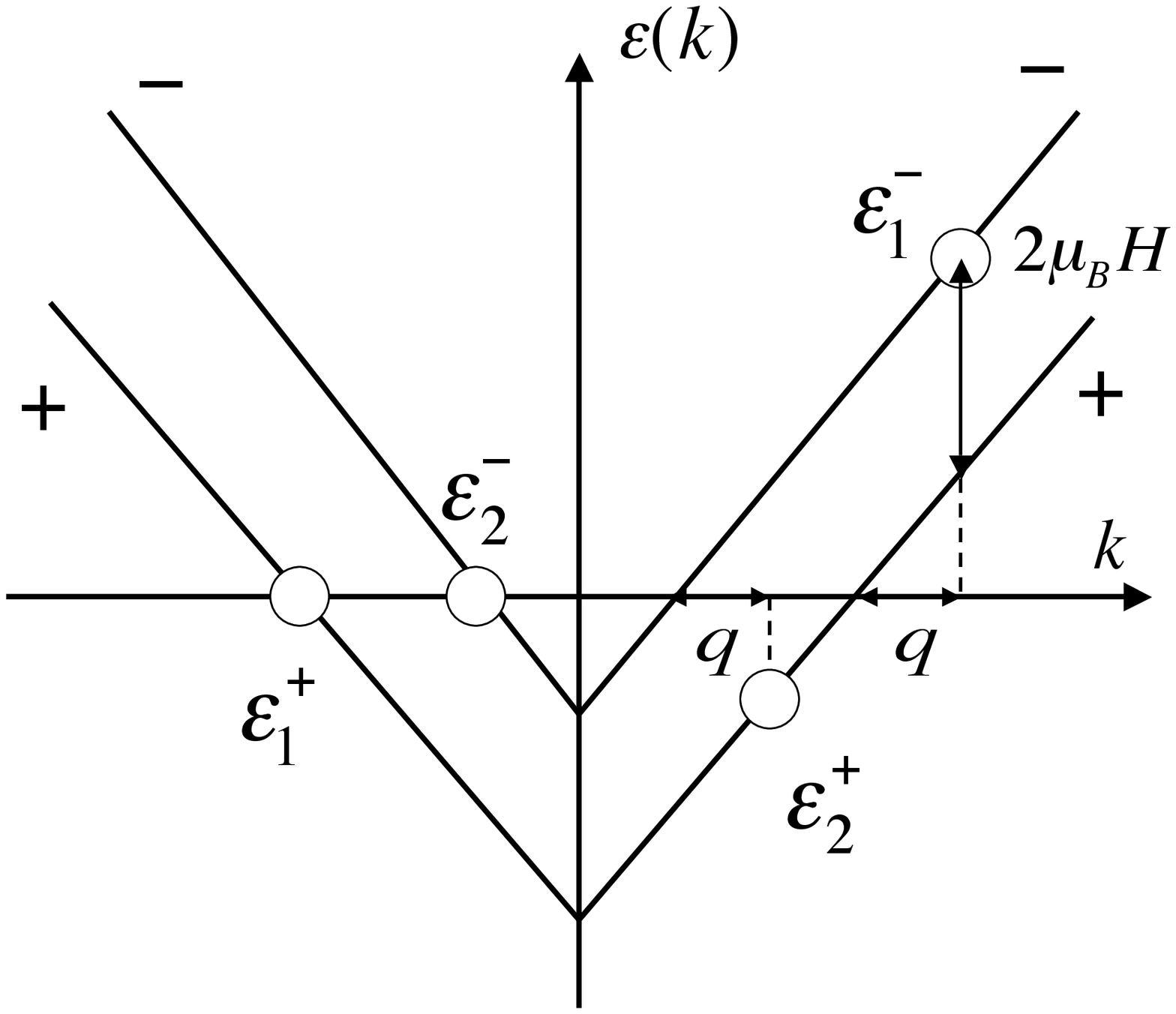}
\caption{ 
Paramagnetic effects split electron spectra with spin up and 
spin down: $\epsilon^{+}(k) = \epsilon_{0}(k) - \mu_B H$ and 
$\epsilon^{-}(k) = \epsilon_{0}(k) + \mu_B H$,
correspondingly.
Two Cooper pairs with spin parts of wave functions,
 $\Delta(+, -)$ and  $ \Delta(-, +)$, and equal total momenta,
 $q  \neq 0$, are characterized by different probabilities to exist
 since energy difference $|\epsilon^{+}_{1} - \epsilon^{-}_{1}|
 = q v_F + 2 \mu_B H$ is not equal to energy difference 
$|\epsilon^{-}_{2} - \epsilon^{+}_{2}|
= - q v_F + 2 \mu_B H$.
[For simplicity, linearized one-dimensional electron spectrum,
$\epsilon (k) = v_F |k|$, is shown].
}
\label{fig1}
\end{figure}

\pagebreak

\begin{table}
\caption{\label{tab:table4} Upper critical fields, $H_{c2}(0)$ [30-33],
transitions temperatures, $T_c$, and triplet-singlet ratio, $R$,
are listed for some modern layered d-wave and s-wave superconductors.
}
\begin{ruledtabular}
\begin{tabular}{ccddd}
&$\beta-(ET)_2AuI_2$&\beta-(ET)_2IBr_2&YBa_2Cu_3O_7&MgB_2\\
\hline
$H_{c2}(0) \ [T]$
  &$5.5 (\parallel)$&2.4 (\parallel)&110  (\perp )&18 (\parallel) \\
  $\mu_B H_{c2} (0) \ [K]$
  &3.7&1.6&74&12\\
  $T_c \ [K]$&4.3&2.3&85 &35\\
  $R$&0.85&0.7&0.85&0.4\\
\end{tabular}
\end{ruledtabular}
\end{table}


\begin{references}
\bibitem{Type-III} We do not use term "type-III superconductivity"
in order to distinguish between suggested in the Letter 
bulk phenomenon and surface superconductivity [1,2].
\bibitem{LebYam} Detailed derivation of Gorkov
equations for singlet-triplet mixed order parameters
and their applications to coexistence of s-wave and p-wave
order parameters will be published elsewhere.
\end{references}
\end{document}